\documentclass[nouppercase]{ifmbe}

\title{Identification of feasible pathway information for c-di-GMP binding proteins in cellulose production}

\affiliation{BioMediTech Institute and Faculty of Biomedical Sciences and Engineering, Tampere University of Technology, Tampere, Finland }{FIRSTAFF}
\affiliation{Laboratory of Chemistry and Bioengineering, Tampere University of Technology, Tampere, Finland }{SECONDAFF}

\author{Syeda Sakira Hassan}{FIRSTAFF}
\author{Rahul Mangayil}{SECONDAFF}
\author{Tommi Aho}{SECONDAFF}
\author{Olli Yli-Harja}{FIRSTAFF}
\author{Matti Karp}{SECONDAFF}

\usepackage{amsthm,amsmath,amssymb}
\usepackage{subcaption}
\usepackage{float}

\begin{document}

\maketitle

\begin{abstract}
In this paper, we utilize a machine learning approach to identify the significant pathways for c-di-GMP signaling proteins. The dataset involves gene counts from $12$ pathways and $5$ essential c-di-GMP binding domains for $1024$ bacterial genomes. Two novel approaches, Least absolute shrinkage and selection operator (Lasso) and Random forests, have been applied for analyzing and modeling the dataset. Both approaches show that bacterial chemotaxis is the most essential pathway for c-di-GMP encoding domains. Though popular for feature selection, the strong regularization of Lasso method fails to associate any pathway to MshE domain. Results from the analysis may help to understand and emphasis to the supporting pathways involved in bacterial cellulose production. These findings demonstrate the need for a chassis to restrict the behavior or functionality by deactivating the selective pathways in cellulose production.
\end{abstract}

\section{Introduction}
The role of cyclic-di-guanosine monophosphate (c-di-GMP) as an allosteric activator for bacterial cellulose synthesis was first discovered by Benziman and coworkers~\cite{Ross1987}. Later, the group identified the genes encoding for enzymes responsible in regulating the c-di-GMP availability in \textit{Komagataeibacter xylinus}. The synthesis and degradation of c-di-GMP are regulated by the catalytic activities of diguanylate cyclases and phosphodiesterases, respectively and identified the presence of similar domain architectures (GGDEF-EAL tandem) among them~\cite{Tal1998}.  Phosphodiesterases containing either EAL or HD-GYP domains involve in c-di-GMP degradation. Genetic and biochemical evidences in several bacterial species demonstrate that the EAL domain containing proteins degrade c-di-GMP to the 5'-phosphoguanylyl-(3'-5')-guanosine (pGpG)~\cite{Bobrov2005}. In contrast to EAL domain, the hydrolyzing activities of c-di-GMP specific phosphodiesterases containing HD-GYP domain result in GMP rather than pGpG. However, biochemical validations are restricted due to unsuccessful purification of catalytically active HD-GYP domain containing proteins. 

Besides the synthesis and degradation of c-di-GMP, proteins that function as c-di-GMP receptors are also important to elicit specific cellular function. Amikam and Galperin reported c-di-GMP binding, PilZ, in the bcsB subunit of \textit{K. xylinus} bacterial cellulose synthase operon. Similar domain (or its homologue) was also identified in other bacterial species such as \textit{Pseudomonas aeruginosa}, \textit{Escherichia coli} and \textit{Vibrio cholera}, involved in c-di-GMP mediated regulation of cellular motility, virulence and biofilm formation~\cite{Pratt2007,Alm1996}. Recently, a new c-di-GMP receptor domain, MshE, was identified in \textit{V. cholarae} and \textit{P. aeruginosa} that contained C-terminal ATP binding site and an N-terminal c-di-GMP binding domain~\cite{Jones2015}. Structural studies report that the c-di-GMP binding affinity of MshE domain was greater than the PilZ domain. 

Taking the importance of c-di-GMP as a universal regulator for several bacterial cellular processes, with the aim to improve bacterial cellulose production, it is rational to study various targets or effectors of c-di-GMP involved in metabolic pathways. Thus, we are interested in finding significant features from the distributions of these c-di-GMP signaling pathways in diverse bacteria.  We propose the use of machine learning approaches that have two advantages. First, we can identify the supporting pathways associated with c-di-GMP signaling proteins. Second, this knowledge can be applied to restrict the behavior of a new strain in synthetic biology. In order to identify relevant features, a predictive model is required that establishes the relationship between the pathways and genes encoding domains. We select in this study two state-of-the-art machine learning approaches, which yield simultaneous predictive models and feature selection. 
\section{Materials and Methods}
\subsection{Data}
For this experiment, we considered $1024$ complete bacterial genomes that are available in the NCBI's RefSeq database. The input data set is downloaded from KEGG, a pathway database that maps for cellular and organismal functions. The output data were downloaded from~\cite{Romling2013}. The selected metabolic pathways and their significance in regulation of c-di-GMP are described in Table \ref{Tab:ListofPathways}. 
\begin{table*}[htb!]
\footnotesize \onehalfspacing
\caption{List of metabolic pathways and their impact on bacterial growth.} 
\centering
\begin{tabular}{lp{11cm}} 
\hline
     \normalfont Metabolic pathways & Significance \\
     \hline
Glycolysis / Gluconeogenesis & Metabolizes glucose to pyruvate. 
 \\
Citrate cycle (TCA cycle) & Generic pathway involving in ATP/GTP production, where c-di-GMP acts as the precursor. \\
Pentose phosphate pathway & Two stage (oxidative and non-oxidative) anabolic pathway generating RNA and aminoacid precursors from glucose. 
\\
Starch and sucrose metabolism & Pathway linking the metabolism of complex carbon substrates such as starch and sucrose to glycolysis route.
\\
Amino sugar and nucleotide sugar metabolism & Pathway involves degradation of amino and nucleotide sugars producing sugar derivatives through glycosylation reaction. 
\\
Lipopolysaccharide biosynthesis & C-di-GMP positively regulates this pathway. The pathway is linked with nucleotide sugar metabolism and pentose phosphate pathway. \\
Sphingolipid metabolism & Pathway involving in the breakdown of lipids that contain sphingoid backbone bases to ceramides. Not commonly present in bacteria.  \\
Terpenoid backbone biosynthesis & Pathway initiates with the condensation of glyceraldehyde 3-phosphate and pyruvate from glycolysis route to produce isoprenoids.   \\
Biosynthesis of amino acids & This pathway is required for cell growth.\\
ABC transporters & Multi-subunit protein family involved in the import (nutrients, trace metals, and vitamins) and export (metabolites) within the bacterial cell.  
\\
Bacterial chemotaxis & Movement of bacteria in response to chemical stimulus. This pathway is regulated by c-di-GMP levels.
\\
Phosphotransferase system (PTS) & Active transport of extracellular substrates into the bacterial cell. \\
\end{tabular}
\label{Tab:ListofPathways}
\end{table*}

\subsection{Regularized Logistic Regression}
Let us consider the observations $(\mathbf{X}, y_i)$ where $\mathbf{X} \in \mathbb{R}^{{n} \times {p}}$ with $n$ observations and $p$ features and $y_i \in \mathbb{R}^{{n} \times {1}}$ in $i^{th}$ domain with $i \in$ \{GGDEF, EAL, HD-GYP, PilZ, MshE\}. A single observation $x_j$ represents a vector of gene counts in the listed pathways and $y_{ij}$ is the number of genes in $i^{th}$ domain for respective bacterial genomes. A linear relationship between $\mathbf{X}$ and $ y_i$ can be modeled as $y_i = \theta_i \mathbf{X}$. 
Here, $\theta_i$ is the relationship parameters which can be estimated by minimizing the residual errors using the equation below.
\begin{equation}\label{eq:leastSquareSol}
	\hat{\theta_i} = \arg \min_{\theta_i}\| y - X\theta_i \|
\end{equation}
where $\|.\|$ is the standard \textit{L$^2$-norm} in the parameter space ~\cite{Stigler1978, Hastie2009}. Although the solution is simple and easily interpretable, it is often inadequate for ill-posed behavior of the underlying data ~\cite{Hastie2009}. In this study, we apply a state-of-the-art regularization approach, \textit{Least absolute shrinkage and selection operator (Lasso)}. The method provides a sparse solution by effectively  shrinking the number of parameters and thereby choosing simpler model~\cite{Tibshirani1996}. In regularization, an extra term, $\lambda$ is added, which controls the trade-off between the residual error and the number of parameters. Thus, our linear model can be defined as $\hat{\theta_i} = \arg \min_{\theta_i}\| y - X\theta_i \| + \lambda\| \theta_i \|_1 $,
where $\lambda > 0$ is the regularization hyperparameter and  $\|.\|_1$ is the \textit{L$^1$-norm} in the parameter space. If we set $\lambda = 0$, it yields to Equation~\ref{eq:leastSquareSol}. On the other hand, a very large $\lambda$ will completely shrinks the parameters to zero and may yield a null or empty model. We use the R package \texttt{glmnet} in this paper~\cite{Friedman2010}.
The model hyperparameter $\lambda$ can be selected using the cross-validation approach \cite{Efron1983}.  In cross validation, the given dataset is randomly divided into training and testing dataset. Training dataset is used for training the model, whereas testing dataset is used for testing the model. The most common variation in cross validation is the $K$-fold cross validation, which is used in this study with $K = 10$.
\subsection{Random Forests}
Random forests is an ensemble learning method where collection of decision trees are built by bootstrap aggregation~\cite{Breiman2001}. A decision tree can be thought of as a hierarchical representation of  if-then rules, where each internal node in the tree describes each input attribute or feature and the leaf node describes the output value. The random forests combines many binary decision trees using several bootstrap samples from the data and choses randomly at each node a subset of the input features. The advantage of random forests is unbiasedness to overlearning, which is done by averaging, thereby improves the prediction accuracy. Since the algorithm ranks the importance of features, it acts as an embedded feature selection approach. In this paper, we use the R package \texttt{randomForest} developed by Liaw and Wiener~\cite{Liaw2002}.
\subsection{Prediction and Feature Selection}
For prediction, we train the Lasso and Random forests models with all the data set except the one for which we measure the quantity of genes encoding different domains. We use the default values of the hyperparameters.

In order to find relevant pathways, we train the models with the complete data set and estimate the importance of the pathways listed in Table~\ref{Tab:ListofPathways}. The regularization technique embedded in Lasso allows removing the irrelevant features from the dataset by setting the coefficient values to zero. For random forest, we use the function \texttt{importance()} to rank the features based on out-of-bag prediction error. Larger values signifies the importance of the features.
\section{Results and Discussion}
\subsection{Results}
\begin{figure}[!htb]
	\centering
    \begin{subfigure}{0.9\columnwidth}
		\includegraphics[width={\textwidth}]{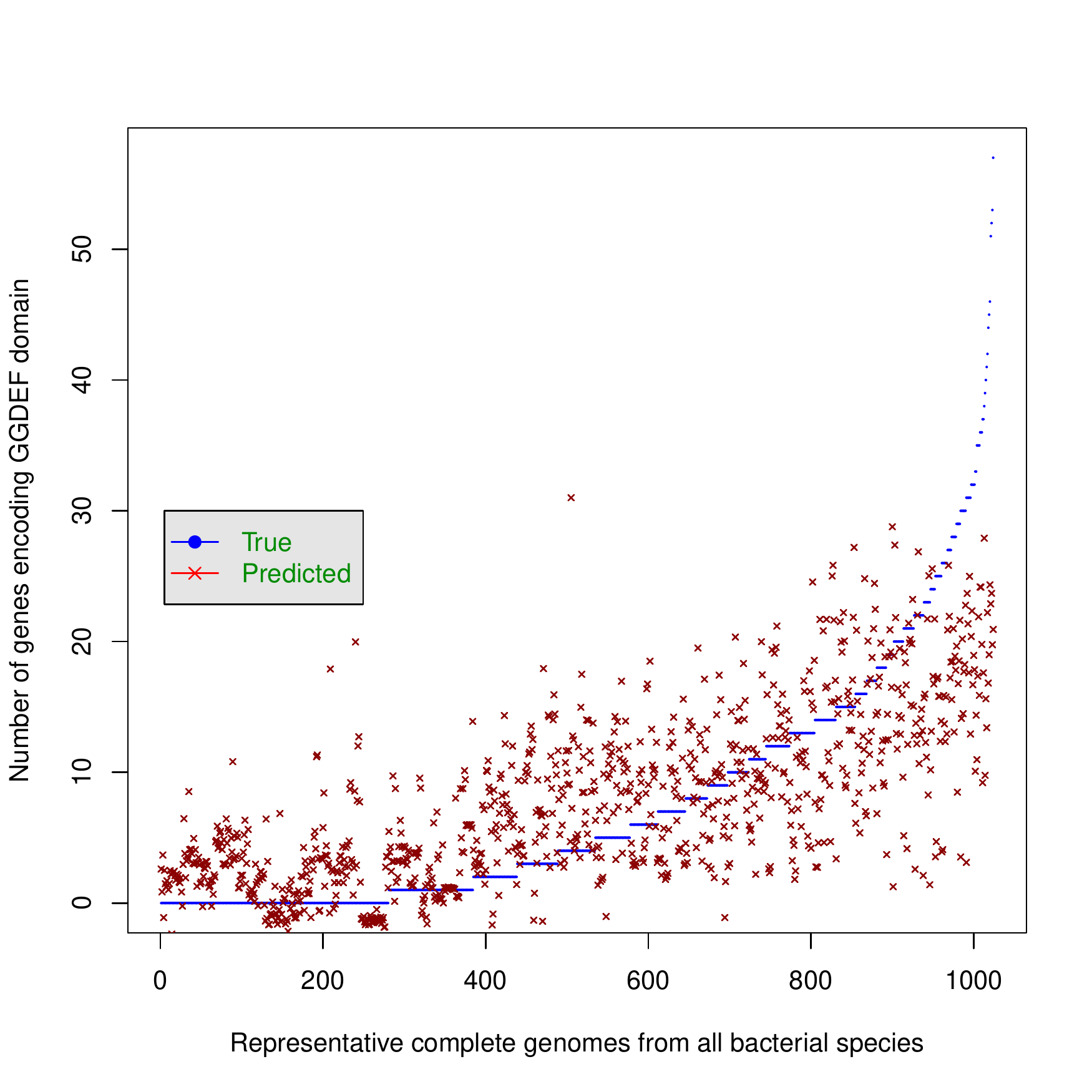}
	\end{subfigure}
    \begin{subfigure}{0.9\columnwidth}
 	\includegraphics[width={\textwidth}]{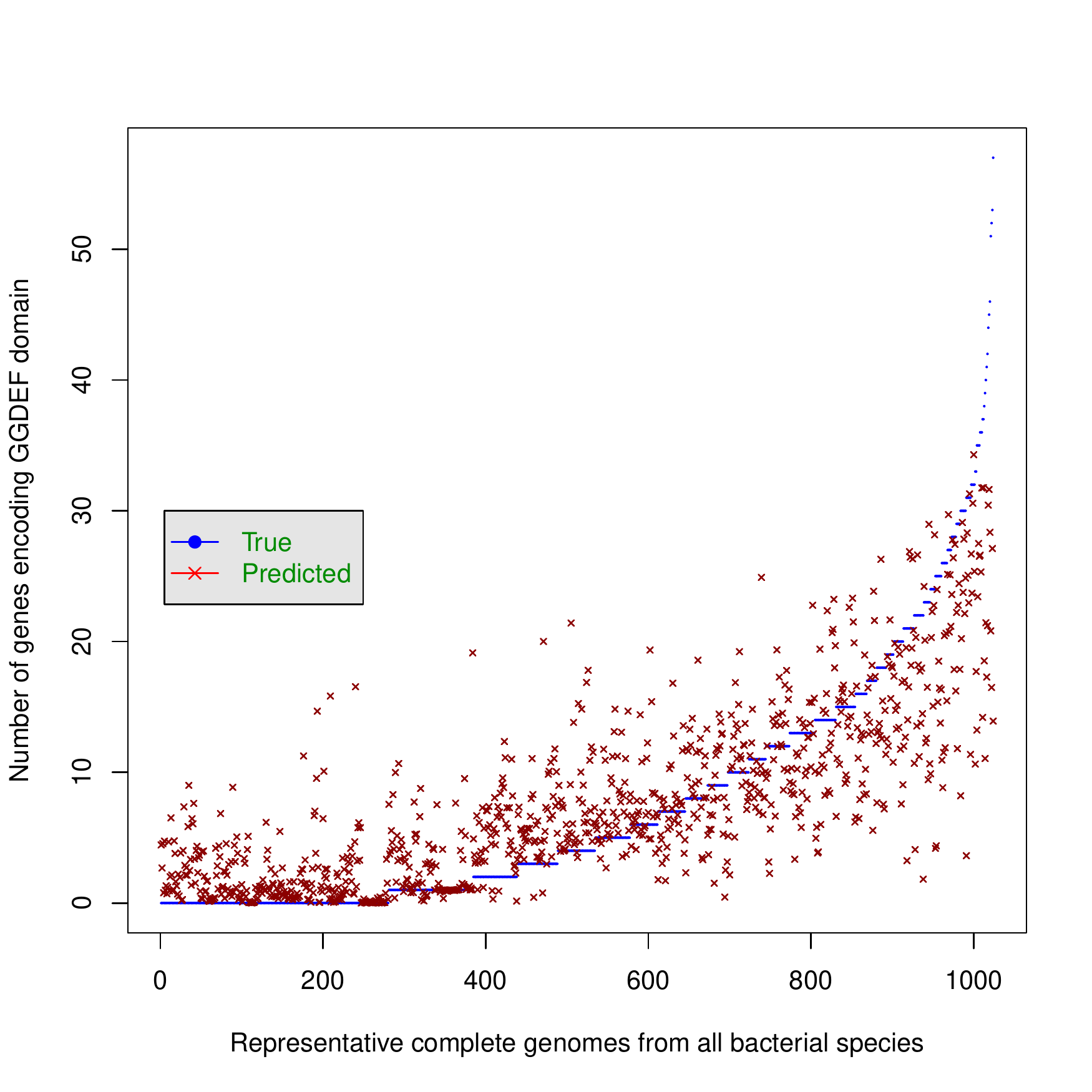}
	\end{subfigure}
\caption{Distribution of genes encoding  c-di-GMP signaling GGDEF domain in respective complete genomes across all bacterial species. Blue circle represents the true distribution and red circle represents the predicted distribution using Lasso (top panel) and random forests (bottom panel) models.}\label{fig:GGDEF}
\end{figure}

In our case study, first we demonstrate the prediction performance of different methods for representative bacterial genomes. Then, we investigate the feature selection approach to find significant pathways associated with c-di-GMP binding proteins in bacterial cellulose production. Figure~\ref{fig:GGDEF} illustrates the true and predictive distribution of genes encoding GGDEF domain for representative bacterial genomes. Here, GGDEF domain is presented as an example. Similar distributions can be drawn for other domains.
\begin{figure}[!htb]
	\centering
		\includegraphics[width=\columnwidth]		{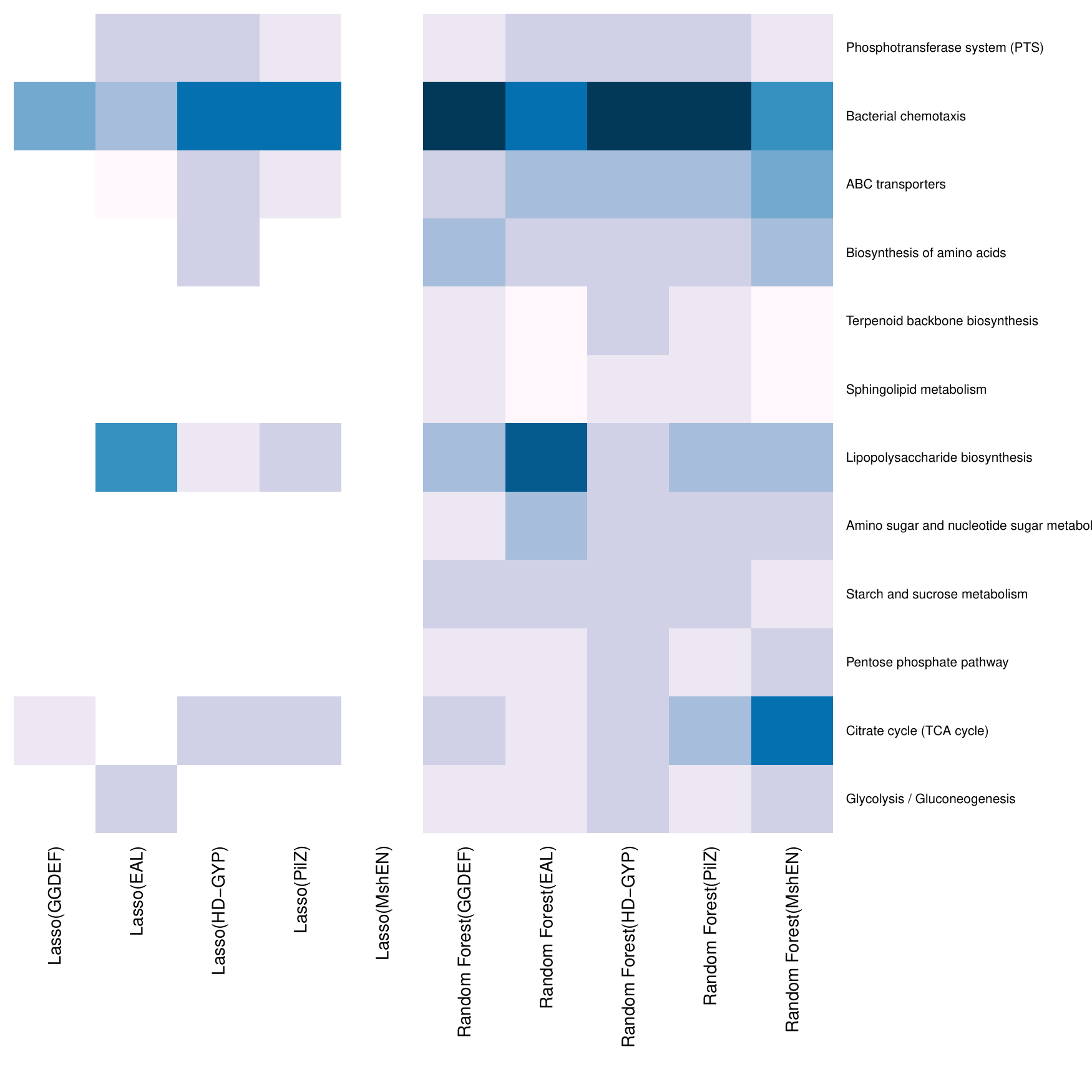}
	\caption{Significance of pathways represented by heatmap.}
    \label{fig:pathways}
\end{figure}

In feature selection, the coefficient values obtained by Lasso and random forests models can be visualized through the heatmap representation in Figure~\ref{fig:pathways}. The higher the coefficient values, the more significant the pathways associated with c-di-GMP signaling domains. In comparison to the metabolic pathways evaluated using Lasso and random forests methods, a denser heatmap is observed for bacterial chemotaxis pathway, indicating its significance for c-di-GMP encoding domains. The lipopolysaccharides biosynthesis is also considered relevant by the methods. The results are analogous to our hypotheses listed in Table~\ref{Tab:ListofPathways}.


\subsection{Discussion}
We examined the feature selection approaches of Lasso and random forests algorithms for c-di-GMP binding proteins. For the first four domains (GGDEF, EAL, HD-GYP, PilZ), the performance of both methods is almost equivalent without requiring any fine-tune for the model hyperparameters. For MshE domain, the random forests method has shown to be competitive with the alternative Lasso approach. According to Lasso, none of the selected pathways is significant for MshE domain. One possible reason is that, the number of genes encoding MshE domain is few. Therefore, the lasso may fail to associate the significance between the pathways and the MshE domain. On the other hand, the nonlinearity inherent in random forests approach facilitates to identify the significance of the pathways.

\section{Conclusion}
The prominent directions in c-di-GMP research has yielded more questions than answers. The addition of machine learning approaches can offer a new insight and understanding in regulation of c-di-GMP binding proteins. We demonstrate in this study only $12$ pathways from KEGG database which is updated constantly. With our approach, it is also possible to integrate and interpret large-scale data set from KEGG database. Identifying relevant metabolic pathways can be an attractive strategy, for example in synthetic biology, by which we can inactivate or silent activate cryptic pathways in bacterial strain for advancement in cellulose production.
\section*{Conflict of Interest}
The authors declare that they have no conflict of interest.
\bibliography{example}
\begin{table}[h]
\footnotesize
        \begin{tabular}{ll}
        &Author: Syeda Sakira Hassan\\
        &Institute: Tampere University of Technology \\
        &Street: Korkeakoulunkatu 10\\
        &City: Tampere\\
        &Country:  Finland\\
        &Email: sakira.hassan@tut.fi\\
        \end{tabular}
\end{table}

\end{document}